\documentstyle[twoside,fleqn,npb,epsfig]{article}

\newcommand{\bea}{\begin{eqnarray}}
\newcommand{\eea}{\end{eqnarray}}

\newcommand{\lsim}{\raisebox{-0.13cm}{~\shortstack{$<$ \\[-0.07cm] $\sim$}}~}

\def\permil{$\%\raise.17ex\hbox{\small$_0$}$}

\newcommand{\AmS}{{\protect\the\textfont2
  A\kern-.1667em\lower.5ex\hbox{M}\kern-.125emS}}

\title{Lepton Flavour Violating $Z$ Decays in the MSSM\thanks{
Talk given at RADCOR and Loops and Legs 2002.}}

\author{Jos\'e I. Illana\address{Centro Andaluz de F{\'\i}sica de 
	Part{\'\i}culas Elementales (CAFPE) and \\
	Departamento de F{\'\i}sica Te\'orica y del Cosmos,
	Universidad de Granada, \\ E-18071 Granada, Spain}
        \thanks{Work supported by CICYT, Junta de Andaluc{\'\i}a and the 
		European Union under contracts FPA2000-1558, FQM-101 and 
		HPRN-CT-2000-00149, respectively.}
	}

\begin{document}

\begin{abstract}
The possibility to observe lepton flavour violating $Z$ decays in the GigaZ 
option of DESY's TESLA project consistently with present bounds from other
processes is analyzed in the context of the minimal supersymmetric standard 
model. In particular, constraints on the slepton mass matrices from radiative 
lepton decays are updated and taken into account.
Their correlation to the present measurement of the muon anomalous dipole
moment is briefly discussed.
\end{abstract}

\maketitle

\section{LEPTON FLAVOUR VIOLATION}

\subsection{Motivation}

Lepton flavour number is perturbatively conserved in the Standard Model (SM)
but a tiny lepton flavour violation (LFV) is expected in charged processes
when including light massive neutrinos ($\nu$SM) compatible with the 
observed neutrino oscillations, {\it e.g.} 
BR$(\ell_J\rightarrow \ell_I \gamma)\lsim 
10^{-48}$ and BR$(Z\rightarrow \ell_I \ell_J)\lsim 10^{-54}$ 
$(\ell_I\ne\ell_J)$ \cite{Illana:2000ic}.
Such effects are far beyond the reach of present or future experiments.
Therefore, the observation of charged LFV would be a clear signature of 
physics beyond the SM.

Supersymmetric (SUSY) models introduce mixings in the sneutrino and the 
charged slepton sectors which could imply flavour--changing processes at 
rates accessible to upcoming experiments. Following a recent work 
\cite{Illana:2002tg}, I report on the TESLA GigaZ 
\cite{Aguilar-Saavedra:2001rg} potential to observe $Z\rightarrow \ell_I 
\ell_J$ in SUSY models with unbroken R--parity, where the slepton mass 
matrices constitute a natural source of LFV. 
At GigaZ the present limits on LFV $Z$ decays from LEP can be improved
by a factor of a hundred to a thousand. An appropriate parameterization 
of the slepton matrices is used. No assumption on the origin of SUSY 
breaking is made. The relevant LFV parameters are allowed to vary consistently
with the present lower limits on SUSY masses and the currently most 
constraning LFV processes: $\mu\to e\gamma$, 
\mbox{$\tau\to e\gamma;\, \mu\gamma$}. 
Others have smaller ratios in SUSY, {\it e.g.}
BR$(\ell_J\rightarrow 3\ell_I)\approx\alpha_{em}$~BR$(\ell_J\rightarrow 
\ell_I \gamma)$ 
or R$(\mu\mbox{Ti}\to e\mbox{Ti})\approx5\times10^{-3}$~BR$(\mu\to e\gamma)$,
giving in both cases weaker bounds, according to present experiments.

For a study of $Z\rightarrow d_I d_J$ in SUSY and two--Higgs--doublet 
models see \cite{Eilam}.

\vspace{-2mm}
\subsection{Properties of the effective vertex}

In order to compare $Z\to\ell_I\ell_J$ and $\ell_J\to\ell_I\gamma$ let us
first examine the Lorentz structure of the effective
$V \bar\ell_I\ell_J$ vertex, since it already reveals some distinctive
features.

The most general vertex $V \bar\ell_I\ell_J$ coupling on--shell 
fermions (leptons) to a vector boson ($V=\gamma$, $Z$) can be parameterized in
terms of four form factors: $F_V$, $F_A$, $F_M$ and $F_E$. 
Unlike the vector and axial--vector (the first two), the magnetic and electric 
dipole form factors are chirality flipping and therefore they are proportional
to a fermion mass, either external or internal (due to virtual fermions 
running in loops).

For an on--shell (massless) photon $F^\gamma_A=0$, and, in addition, if
$m_{\ell_I}\ne m_{\ell_J}$ then $F^\gamma_V=0$. As a consequence, the 
flavour--changing process $\ell_J \rightarrow \ell_I \gamma$
is determined by (chiraliy--flipping) dipole transitions only. 
In contrast, all form factors and chiralities contribute to the decay of a 
$Z$ boson.

The anomalous magnetic dipole moment is related to $F_M^\gamma$ for identical 
external leptons.

\section{LFV IN SUPERSYMMETRY}

\subsection{SUSY contributions}

\begin{figure*}
\centerline{\epsfig{file=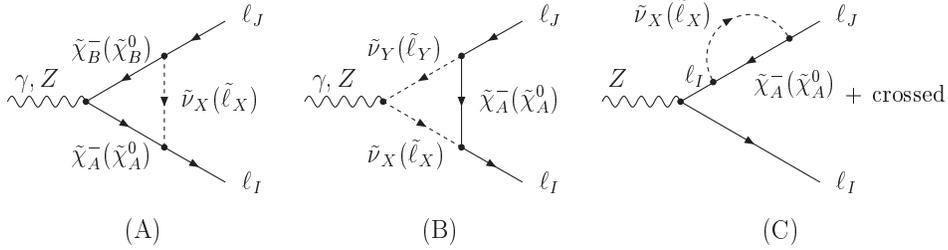,width=0.8\linewidth}}
\caption{SUSY contributions to the LFV processes $Z\to\ell_I\ell_J$ and
$\ell_J\to\ell_I\gamma$.
\label{fig1}}
\end{figure*}

The genuine supersymmetric contributions to one loop to the LFV 
processes $Z\to\ell_I\ell_J$ and $\ell_J\to\ell_I\gamma$ are summarized in 
Fig.~\ref{fig1}. There is no contribution at tree level.
Note that chargino--sneutrino diagrams in (A) and
neutralino--charged leptons in (B) do not couple to the photon. Diagrams
of type (C) are not relevant to the photon processes either, since they do 
not give dipole contributions.

It is remarkable that only the sum of all 
chargino--sneutrino diagrams on one side and neutralino--charged sleptons
on the other are ultraviolet finite
and, at the same time, exhibit the decoupling of heavy SUSY particles
running in the loops. This helps as a crosscheck for the calculation. 

To get LFV two conditions have to be fulfilled: sleptons must mix and
their spectrum must be non--degenerate.

\subsection{Slepton mass matrices}

Since SUSY is broken, fermion and scalar mass matrices will be diagonalized 
by different rotations in flavour space. This supplies new sources of LFV.

We assume that the mixing takes place only between two generations ($IJ$),
neglecting the slepton mixing with a third family. Furthermore it is natural
to assume no alignment between fermion and scalar fields, which implies a
mixing angle of order one. Only the slepton mass 
splitting is then left as a free LFV parameter. The different contributions
are separated in LL, RR or LR, according to the entries of the 
slepton mass matrix involved the splitting (only LL for sneutrinos).
The relevant entries of the (symmetric) mass matrices can be written as
\bea
{\bf M}^2_{\tilde\nu}&=&{{\tilde m}}^2\left(\begin{array}{cc} 1 & \cdot \\ 
{\delta^{\tilde\nu\; IJ}_{LL}} & 1
\end{array}\right)\;, \\
{\bf M}^2_{\tilde\ell}&=&{{\tilde m}}^2\left(\begin{array}{cccc}
1 & \cdot & \cdot & \cdot \\
{\delta^{\tilde\ell\; IJ}_{LL}} & 1 & \cdot & \cdot \\
\delta^{\tilde\ell\; II}_{LR} & {\delta^{\tilde\ell\; IJ}_{LR}} & 1 & 
								\cdot \\
{\delta^{\tilde\ell\; JI}_{LR}} & \delta^{\tilde\ell\; JJ}_{LR} & 
{\delta^{\tilde\ell\; IJ}_{RR}} & 1 \end{array}\right)\;,
\eea
where only one $\delta^{IJ}$ is taken different from zero in each case.
Note that $\delta^{\tilde\ell\; II}_{LR}$ and $\delta^{\tilde\ell\; JJ}_{LR}$
are flavour conserving. There are two free parameters:
the mass eigenvalues $\tilde m_1^2$ and $\tilde m_2^2$ or else the mass scale 
$\tilde m^2=\tilde m_1 \tilde m_2$ and the mass splitting 
\mbox{$\delta=(\tilde m_2^2-\tilde m_1^2)/(2\tilde m^2)$}:
\bea
\tilde m_{1,2}^2=\tilde m^2 (\sqrt{1+\delta^2}\mp \delta)\;.
\eea
The splitting is responsible for any flavour--changing process: 
$\delta=0$ corresponds to the flavour--conserving case, $\delta\ll 1$
can be treated as a non--diagonal mass insertion, and 
$\delta\rightarrow \infty$
gives $\tilde m_2^2\rightarrow \infty$ (a decoupled second family).
The last case implies a maximum flavour--changing rate.

\subsection{Calculation}

We have obtained analytical expressions for the amplitudes of the processes
under study in terms of generic couplings and one--loop tensor integrals
\cite{Illana:2002tg}. The complete set of Feynman rules, including 
full chargino and neutralino mixings, have been implemented to get
the numerical results. 
We have taken into account the direct constraints on 
SUSY masses in Table~\ref{tab1}. Note that, for negligible scalar trilinears
$m_{\tilde\nu}^2=m_{\tilde\ell_L}^2+M^2_Zc^2_W\cos2\beta$,
and the bounds on $m_{\tilde\nu}$ and $m_{\tilde\ell_L}$ are 
correlated. For instance:
$m_{\tilde\nu}> 65\ (40)\ \mbox{GeV for}\ \tan\beta=2\ (50)$.

\begin{table}
\caption{Approximate lower bounds on SUSY mass parameters based on 
\cite{Groom:in}.
\label{tab1}}
\begin{tabular}{ll}
\hline
sleptons $(\tilde m_1)$  & $m_{\tilde\nu} > 45$~GeV \\
 				 & $m_{\tilde\ell_{L,R}} > 90$~GeV
\\
\hline
charginos & $m_{\tilde\chi^+_1} > 75$~GeV, 
					if $m_{\tilde\nu}>m_{\chi^+_1}$ \\
                   & $m_{\tilde\chi^+_1} > 45$~GeV, otherwise 
\\
\hline
neutralinos & $m_{\tilde\chi^0_1} > 35\mbox{ GeV}$
\\
\hline
\end{tabular}
\end{table}

There are five input parameters in the SUSY sector to be scanned consistently
with above bounds: $M_2$, $\mu$ and $\tan\beta$ (controlling the spectra and
couplings of charginos and neutralinos), $\tilde m_1$ (relevant lightest
scalar mass) and $\delta$ (LFV parameter).

\begin{figure}
\epsfig{file=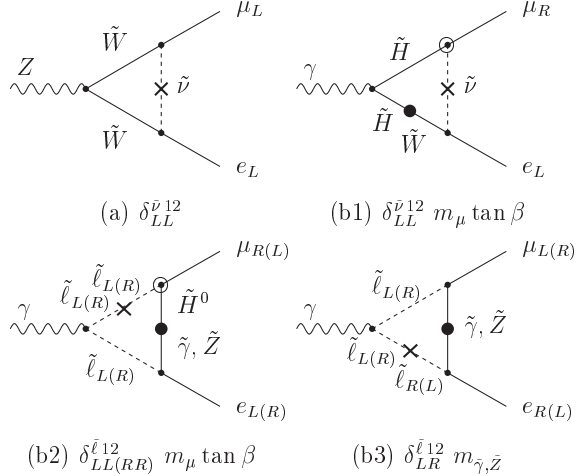,width=\linewidth}
\caption{Dominating diagrams contributing to (a) $Z\to\mu e$ and
(b) $\mu\to e\gamma$, showing the approximate linear dependence on the 
$\delta$ insertions (crosses), the fermion mass insertions (big dots) and
the Yukawa couplings (open circles). \label{fig2}}
\end{figure}

It is instructive to examine the diagrams in terms of 
gauginos, current eigenstates and mass insertions (the amplitudes
are approximately proportional to $\delta$ when this is small). The dominating
diagrams are depicted in Fig.~\ref{fig2}. The lepton chirality is
specified. All the diagrams contributing to $\ell_J\to\ell_I\gamma$ are
proportional to $\tan\beta$ times an external fermion mass (because they
pick a Yukawa coupling), except for the last one that is proportional
to a gaugino mass.
Due to the weaker experimental bounds on sneutrino masses, 
the dominant contributions to $Z\rightarrow \ell_I \ell_J$ come
from the diagrams mediated by charginos and sneutrinos. 

It is important to emphasize that our results depend on contributions 
with opposite signs that often cancel when varying a parameter. For example, 
one would expect that the branching ratios are optimized for light slepton 
masses. However, we observe frequently the opposite effect. The rates 
can increase by raising the mass of the sleptons up to values of 500 
GeV, and only at masses above $1-2$ TeV the asymptotic regime is reached 
\cite{Illana:2002tg}. 

\section{SUSY PREDICTIONS FOR $Z\to\ell_I\ell_J$}

Let us first consider the process $Z\rightarrow \ell_I \ell_J$ uncorrelated
from other LFV processes. For SUSY masses
above the current limits it is possible to have 
$Z\rightarrow \mu e ; \tau e; \tau\mu$ 
at the reach of GigaZ.
The maximun rate is obtained when the second slepton, 
$\tilde\ell_J$, is very heavy \mbox{($\delta^{IJ}\rightarrow\infty$)}. 
The largest contribution
comes from virtual sneutrino--chargino diagrams (all other contributions
are at least one order of magnitude smaller). It gives  
${\rm BR}(Z\rightarrow \ell_I \ell_J)$ from $2.5\times 10^{-8}$ for 
$\tan\beta=2$ to $7.5\times 10^{-8}$ for $\tan\beta=50$, practically 
independent of the lepton masses. The variation
is due to the mild dependence of chargino and
sneutrino masses on $\tan \beta$. We find that a branching ratio larger than
$2\times 10^{-9}$ ($2\times 10^{-8}$) can be obtained with sneutrino
masses of up to 305 GeV (85 GeV) and chargino masses of up to
270 GeV (105 GeV). These branching ratios are well below
LEP limits of ${\cal O}(10^{-6})$ but range within GigaZ sensitivities.

However, most of these values of ${\rm BR}(Z\rightarrow \ell_I \ell_J)$
are correlated with an experimentally excluded rate of 
$\ell_J\rightarrow \ell_I \gamma$. In particular, after
scanning for all the parameters in the model we find that 
${\rm BR}(\mu\rightarrow e \gamma)<1.2\times 10^{-11}$ implies
\mbox{${\rm BR}(Z\rightarrow \mu e)<1.5\times 10^{-10}$},  
below GigaZ reach.

\begin{figure*}
\begin{center}
\begin{tabular}{ccc}
\epsfig{file=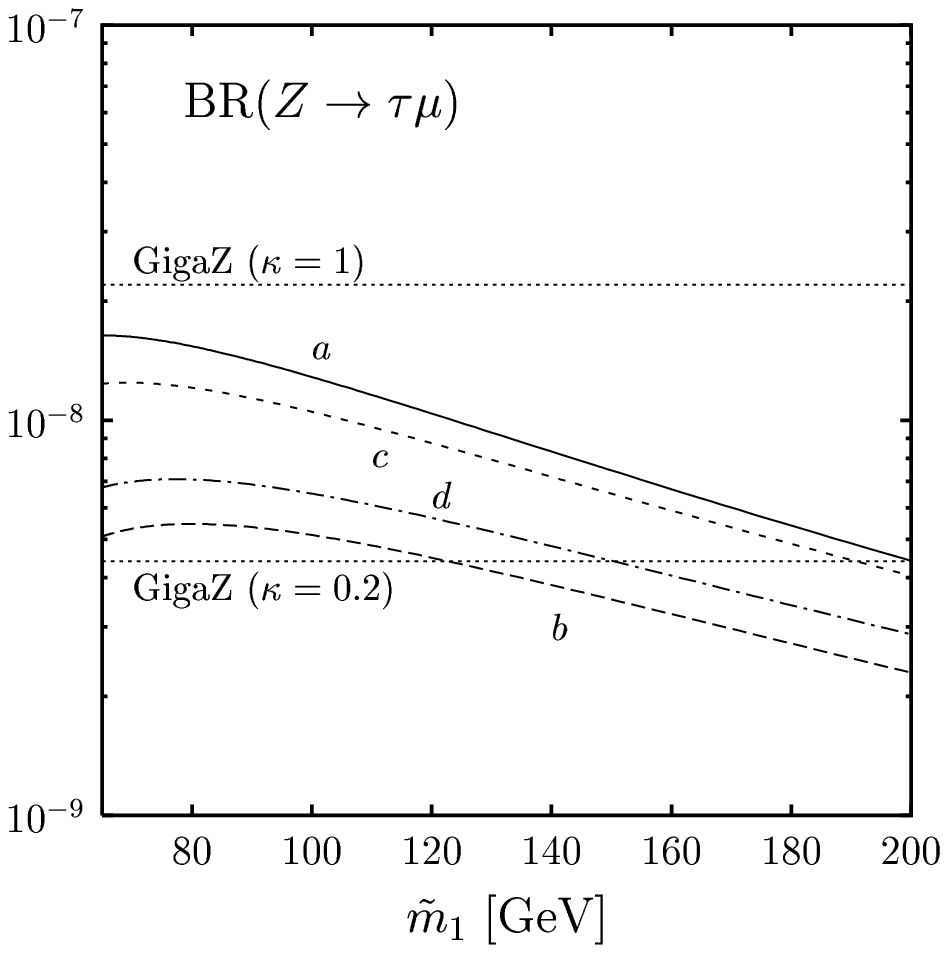,width=0.4\linewidth} &\hspace*{1cm}&
\epsfig{file=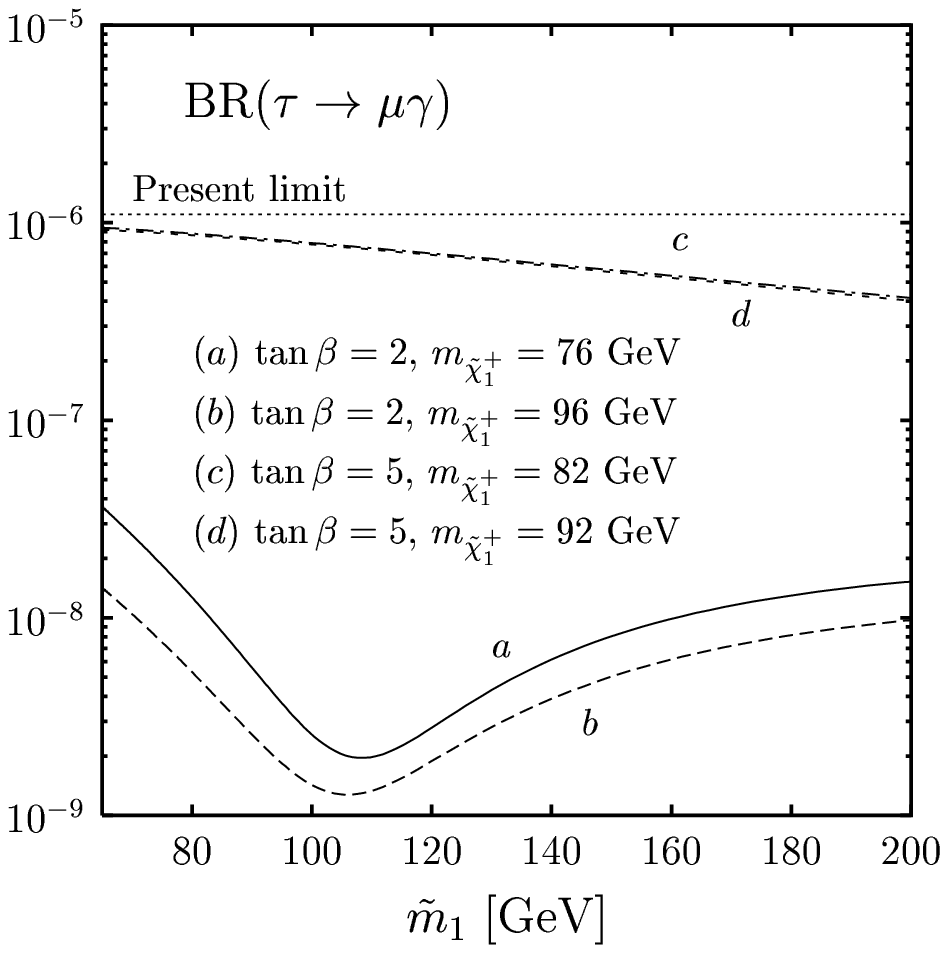,width=0.4\linewidth}
\end{tabular}
\end{center}
\vspace{-10mm}
\caption{BR$(Z\to\tau\mu)$ and $BR(\tau\to\mu\gamma)$
 as a function of the lightest sneutrino mass ($\tilde m_1$) with the
other one decoupled ($\delta^{\tilde\nu\,23}_{LL}\to\infty$), in several 
SUSY scenarios at the reach of the best GigaZ projection ($\kappa=2$).
\label{fig3}}
\end{figure*}

A more promising result is obtained for the processes involving
the $\tau$ lepton. It turns out (see next Section) that 
the bounds from $\tau\rightarrow e\gamma; \mu\gamma$ can be
avoided while still keeping 
$Z\rightarrow \tau e;\tau\mu$ at the reach of the best GigaZ
projection (see Fig.~\ref{fig3}). In particular, for large 
$\delta^{\tilde\nu\,13,23}_{LL}$ and a light sneutrino we get 
${\rm BR}(Z\rightarrow \tau e)\approx 
{\rm BR}(Z\rightarrow \tau\mu)\approx 1.6\times 10^{-8}$
for ${\rm BR}(\tau\rightarrow e \gamma)\approx 
{\rm BR}(\tau\rightarrow \mu \gamma)\approx 3.5\times 10^{-8}$
(two orders of magnitude below current limits!).
We obtain events at the reach of GigaZ with lightest sneutrino 
masses from 55 to 215 GeV, lightest chargino from 75 to 100 GeV, and 
$\tan\beta$ up to 7. They all need large and negative values of the 
Higgsino mass parameter $\mu$.
The contributions due to charged slepton mixing saturate the experimental 
bounds to $\tau\rightarrow e\gamma;\mu \gamma$ giving at most an effect  
one order of magnitude below the reach of GigaZ in 
$Z\rightarrow \tau e;\tau\mu$.

\section{BOUNDS FROM $\ell_J\to\ell_I\gamma$}

The bounds on the mass insertions $\delta^{IJ}$ 
establish how severe is the flavour problem in the
lepton sector of the MSSM, since they provide the
most stringent constraints on the slepton LFV mass
terms available today. 
We have updated the early works by Masiero and
collaborators \cite{Masiero} to include more recent
experiments and, in particular, the complete calculation
in the MSSM: only photino--mediated diagrams had
been considered in \cite{Masiero}, but they are typically subdominant
as pointed out already by Ref.~\cite{Feng:2001sq}. Moreover, the
frequent cancellations of different contributions demand a careful 
treatment.

To estimate the MSSM prediction we combine low and 
high values of the relevant parameters: $\tan\beta=2;50$,
$\tilde m_1=100;500$ GeV, and the gaugino and higgsino mass 
parameters $M_2=150;500$ GeV and $\mu=\pm 150;\pm 500$ GeV. A
summary of the results \cite{Illana:2002tg} follows. 

The bounds on the first two families are very restrictive. 
For $\delta^{\tilde\nu\,12}_{LL}$, $\delta^{\tilde \ell\,12}_{LL}$
and $\delta^{\tilde \ell\,12}_{RR}$ they range from ${\cal O}(10^{-3})$
to ${\cal O}(10^{-5})$ and are stronger for high $\tan\beta$.
This demands a very high degeneracy between the selectron and the
smuon.
For $\delta^{\tilde \ell\,12}_{LR}$ the limits are of ${\cal O}(10^{-6})$,
practically independent of $\tan\beta$. This implies just that the
scalar trilinears, usually assumed proportional to Yukawa couplings,
are small.

The experimental bounds on the mass insertions involving the 
third family are much weaker. In particular, for small $\tan\beta$
we find no bounds on any $\delta^{I3}$ (except for 
$\delta^{\tilde\ell\,I3}_{LR}$). For large $\tan\beta$ the
bounds are (depending on the values of the SUSY--breaking masses)
$\delta^{\tilde \nu\,I3}_{LL} = 0.03 \;{\rm to}\; 1.3$; 
$\delta^{\tilde \ell\,I3}_{LL} = 0.14 \;{\rm to}\; \infty$; 
and 
$\delta^{\tilde \ell\,I3}_{RR} = 0.11 \;{\rm to}\; \infty$.
For the $LR$ mass insertions we find
$\delta^{\tilde \ell\,I3}_{LR} = 0.05 \;{\rm to}\; \infty$,
independent of $\tan\beta$.

\section{RELATION TO $(g_\mu-2)$}

A $g_\mu-2$ correction would be generated by the diagrams
in Fig.~\ref{fig2}b if no mass insertions $\delta^{\tilde \nu\,IJ}_{LL}$,
$\delta^{\tilde \ell\,IJ}_{LL}$, $\delta^{\tilde \ell\,IJ}_{RR}$ are 
included and $\delta^{\tilde \ell\,IJ}_{LR}$ is replaced by 
$\delta^{\tilde \ell\,22}_{LR}$.
In this sense, $g_\mu-2$ is a normalization of the
branching ratio BR$(\ell_J\rightarrow \ell_I \gamma)$ for processes changing
the muon flavour.

The new data from the Brookhaven muon $g-2$ experiment \cite{Bennett:2002jb}
confirms the previous measurement with twice the precision.
Since the (revised) SM prediction \cite{Knecht:2001qf} was already off by
more than one standard deviation, the present `discrepancy' has increased 
and is now approximately 
$\delta a_\mu=a^{\rm exp}_\mu-a^{\rm SM}_\mu=(24\pm10)\times 10^{-10}$.
It is nevertheless affected by significant theoretical uncertainties, since
different groups disagree in their estimate of the hadronic contributions.

Trying to extract conclusions of possible new physics is therefore
rather speculative but, taking the discrepancy seriously, it seems to 
indicate that the muon dipole moment may need non--standard contributions 
of positive sign. Blaming it on SUSY,
we obtain, in agreement with \cite{Moroi:1996yh}, positive or negative 
contributions correlated with the sign of the Higgsino mass parameter 
$\mu$ and similar in size to the weak corrections. 
Extra assumptions have to be made in order to constrain LFV processes from
$(g_\mu-2)$ but, in any case, the favourite region for $Z\to\tau\mu$ at
GigaZ, requiring a negative $\mu$ parameter (Fig.~\ref{fig3}), 
seems disfavoured.

\section{CONCLUSIONS}

SUSY models introduce LFV corrections which are proportional to 
slepton mass squared differences. We have shown that the non--observation
of $\mu\to e\gamma$ precludes the observation of $Z\to\mu e$ at GigaZ
and implies at least a one permil degeneracy between the lightest slepton
families (maybe justifiable by the weakness of their Yukawa couplings).
In contrast, the current bounds on $\tau\to e\gamma;\,\mu\gamma$ introduce
only weak constraints (no flavour problem for the third lepton family) and
there is a (small) window for the observation of $Z\to\tau\mu$ at the best
projection of GigaZ.

\vspace{1mm}
\noindent
{\it Acknowledgements}.
It is a pleasure to thank Manuel Masip for a fruitful collaboration
and the organizers of this meeting for their willingness and the pleasant 
atmosphere of the event.


\begin{thebibliography}{9}

\bibitem{Illana:2000ic}
J.~I.~Illana and T.~Riemann,
Phys.\ Rev.\ D {\bf 63} (2001) 053004;
J.~I.~Illana, M.~Jack and T.~Riemann,
hep-ph/0001273.

\bibitem{Illana:2002tg}
J.~I.~Illana and M.~Masip,
hep-ph/0207328.

\bibitem{Aguilar-Saavedra:2001rg}
J.~A.~Aguilar-Saavedra {\it et al.}  [ECFA/DESY LC Physics Working Group
                  Collaboration],
hep-ph/0106315.

\bibitem{Eilam}
G.~Eilam, these Proceedings.

\bibitem{Groom:in}
D.~E.~Groom {\it et al.}  [Particle Data Group Collaboration],
Eur.\ Phys.\ J.\ C {\bf 15} (2000) 1.

\bibitem{Masiero}
F.~Borzumati and A.~Masiero,
Phys.\ Rev.\ Lett.\  {\bf 57} (1986) 961;
F.~Gabbiani and A.~Masiero,
Nucl.\ Phys.\ B {\bf 322} (1989) 235;
F.~Gabbiani, E.~Gabrielli, A.~Masiero and L.~Silvestrini,
Nucl.\ Phys.\ B {\bf 477} (1996) 321.

\bibitem{Feng:2001sq}
J.~L.~Feng, K.~T.~Matchev and Y.~Shadmi,
Nucl.\ Phys.\ B {\bf 613} (2001) 366.

\bibitem{Bennett:2002jb}
G.~W.~Bennett {\it et al.}  [Muon g-2 Collaboration],
Phys.\ Rev.\ Lett.\  {\bf 89} (2002) 101804
[Erratum-ibid.\  {\bf 89} (2002) 129903].

\bibitem{Knecht:2001qf}
M.~Knecht and A.~Nyffeler,
Phys.\ Rev.\ D {\bf 65} (2002) 073034;
M.~Knecht, A.~Nyffeler, M.~Perrottet and E.~De Rafael,
Phys.\ Rev.\ Lett.\  {\bf 88} (2002) 071802;
M.~Hayakawa and T.~Kinoshita,
hep-ph/0112102;
I.~Blokland, A.~Czarnecki and K.~Melnikov,
Phys.\ Rev.\ Lett.\  {\bf 88} (2002) 071803;
J.~Bijnens, E.~Pallante and J.~Prades,
Nucl.\ Phys.\ B {\bf 626} (2002) 410.

\bibitem{Moroi:1996yh}
T.~Moroi,
Phys.\ Rev.\ D {\bf 53} (1996) 6565
[Erratum-ibid.\ D {\bf 56} (1996) 4424].

\end{thebibliography}
\end{document}